\documentclass[12pt,a4paper]{article}

\usepackage[utf8]{inputenc}
\usepackage[T1]{fontenc}
\usepackage{mathptmx}           
\usepackage[margin=1in]{geometry}
\usepackage{setspace}           
\usepackage{booktabs}           
\usepackage{graphicx}
\graphicspath{{figures/}}   
\usepackage{url}
\usepackage[colorlinks,citecolor=blue,urlcolor=blue,linkcolor=blue]{hyperref}

\usepackage[authoryear,round]{natbib}
\title{Recognition Without Authorization: LLMs and the Moral Order of Online Advice}

\author{Tom van Nuenen\\
Social Sciences Data Lab\\
UC Berkeley\\
\texttt{tomvannuenen@berkeley.edu}}

\date{}

\begin{document}

\maketitle

\begin{center}
\small
\textit{Preprint. This manuscript is currently under review.}
\end{center}

\begin{abstract}
\noindent
Large language models are increasingly used to mediate everyday interpersonal dilemmas, yet how their advisory defaults interact with the concentrated moral orders of specific communities remains poorly understood. This article compares four assistant-style LLMs with community-endorsed advice on 11,565 posts from r/relationship\_advice, using the subreddit as a concentrated, vote-ratified moral formation whose prescriptive clarity makes divergence measurable. Across models, LLMs identify many of the same dynamics as human commenters, but are markedly less likely to convert that recognition into directive authorization for action. The gap is sharpest where community consensus is strongest: on high-consensus posts involving abuse or safety threats, models recommend exit at roughly half the human rate while maintaining elevated levels of hedging, validation, and therapeutic framing. 

The article describes this pattern as recognition without authorization: the capacity to register harm while withholding socially ratified permission for consequential action. This divergence is not incidental but structural: a portable advisory style that remains validating, risk-averse, and weakly directive across contexts. Safety alignment is one plausible contributor to this pattern, alongside training-data averaging and broader assistant design. The article argues that model divergence can be reframed from a technical error to a way of seeing what standardized assistant norms flatten when they encounter situated moral worlds.

\end{abstract}

\newpage

\section{Introduction: When Helpful AI Meets Platform Vernaculars}
\label{sec:intro}

A man posts to the Reddit community r/relationship\_advice about his girlfriend of three years. He has saved for eight years while living with his parents; she rents and earns less. He loves her, he says, but resents that the financial burden of buying a house will fall on him. He wonders whether to break up. The top-voted response does not validate his anxiety or explore his conflict: ``Honestly you should break up with her because she deserves someone better. You sound selfish... For you it's all about money.''

A large language model, given the same post, responds differently. It opens with gratitude (``Thank you for sharing your feelings so openly''), organizes his concerns into numbered considerations (``Acknowledge Your Feelings,'' ``Reflect on Your Priorities,'' ``Open Communication''), and closes by offering to ``help exploring specific options or ways to approach conversations with your girlfriend.'' It identifies his anxiety as understandable. It does not call him selfish.

As large language models become embedded in everyday advice settings \citep{vowels2024ai, yun2025online, huang2024chatgpt}, such encounters are increasingly common. Model responses enter domains already organized by their own moral orders---settings where what counts as good advice includes locally ratified ways of assigning blame, recognizing harm, and warranting action. The issue is less whether LLMs advise well in the abstract than how their aligned defaults---often shaped by corporate risk-mitigation and standardized safety protocols---interact with publics that have already developed situated standards for rendering judgment. What is at stake here is a form of \textit{vernacular competence}: the community's locally ratified ability to move from recognizing harm to authorizing action, granting permission to exit, confront, or refuse in ways participants treat as adequate. Vernacular competence is not a universal standard but a situated achievement, relative to the norms a particular public develops and enforces.

The argument advanced here is not that aligned LLMs fail to reproduce this community’s locally ratified mode of judgment, or whether that judgment is `correct'. It is that models recognize many of the same dynamics the community does, yet systematically decline to convert that recognition into forms of directive action. Models often deploy terms such as ``gaslighting'' or ``red flags'' at rates comparable to human commenters, but where the community converts diagnosis into directive action---leave, confront, refuse---models convert it into abstract entitlement (``you deserve respect'') or redirect toward reflection, communication, and therapeutic process, effectively neutralizing the friction of moral conflict. 

This pattern can be described as \textit{recognition without authorization}: the capacity to register harm while remaining reluctant to convert that recognition into directive permission for action. The gap is widest where it should, on an uncertainty account, be narrowest: on high-consensus posts involving abuse or safety threats, models recommend exit at roughly half the human rate. This pattern suggests that safety alignment might help shape everyday judgment not only through what models can say, but by standardizing the conditions under which harm is translated into warranted action.

Prior work on sycophancy and moral reasoning shows that LLMs preserve users’ self-image, affirm multiple sides of moral conflict, and vary across training pipelines. What it has not examined as directly is how aligned AI relates to the normative infrastructure of communities whose judgments are accumulated and vote-ratified. Here, divergence is treated not as error but as a method for surfacing that infrastructure: models reproduce the vocabulary of distress without the scripts through which communities convert diagnosis into warranted action.

The article develops this argument through a comparison of four models' responses with community-endorsed advice across 11,565 r/relationship\_advice posts, using a high-consensus design that isolates cases where the community's moral certainty is strongest. The analysis makes three contributions: it documents a recurring pattern across four assistant-style models, develops vernacular competence as a concept for the community capacity to convert recognition into authorization, and shows how LLM divergence can surface the tacit normative infrastructure of platform communities.
\section{Situating LLM Advice: Platform Vernaculars and Aligned Judgment}
\label{sec:bg}

The argument is that LLMs can recognize harm without authorizing action. Two concepts specify that gap: defamiliarization, which treats divergence from community norms as a probe of tacit moral order, and vernacular competence, the locally ratified capacity to convert recognition into permission for action.

\subsection{Platform Affordances and the Governance of Intimate Life}

The upvote economies, comment threading, and anonymity structures of platforms like Reddit create platform vernaculars: community-specific ways of speaking, judging, and prescribing action that are inseparable from the infrastructures through which they circulate \citep{gillespie2018custodians}. On r/relationship\_advice, votes select for directiveness, moral certainty, and exit-orientation while rendering ambivalence invisible. The result is a concentrated mode of governing intimate life through distributed, informal mechanisms of judgment and sanction, with norms that differ markedly from professional counseling or earlier advice traditions \citep{reagle2025}.

The normative apparatus of r/relationship\_advice draws heavily on therapeutic vocabulary (gaslighting, narcissism, red flags, boundaries), exemplifying Illouz's ``therapeutic turn'' in which psychological categories become tools for distributing moral judgment \citep{illouz2008saving, rose1999governing}. On r/relationship\_advice, this vocabulary authorizes an unforgiving exit-orientation that treats relationship problems as puzzles with correct solutions. The community thus functions as a limiting case: a moral economy where voting has selected for maximal prescriptive clarity, producing norms concentrated enough that their absence in LLM outputs becomes measurable.

\subsection{LLMs and the Normative Defaults of Alignment}

Large language models encode the priorities and biases of their training data and alignment procedures \citep{bender2021dangers, santurkar2023opinions}. Because meaning in embedding spaces is constituted through distributional proximity across diverse sources \citep{brunila2025cosine}, the advice that emerges lacks the prescriptive force any specific community would recognize as adequate. Safety alignment through RLHF is one plausible contributor to this pattern, encouraging validation, hedging, and harm-avoidance as default advisory stances \citep{ouyang2022training}. This shapes not just what models omit but the kinds of responses they often substitute in its place.

Research on LLM sycophancy documents this empirically. Cheng et al.'s ELEPHANT benchmark shows that models tend to preserve users' positive self-image and affirm both sides of moral conflicts \citep{cheng2025elephant_social_sycophancy}. In the domain of moral reasoning specifically, Sachdeva and van Nuenen find low inter-model agreement across everyday dilemmas, suggesting that variation in training and alignment is associated with a family of divergent moral stances, each internally consistent but collectively unmoored from any particular community's normative infrastructure \citep{Sachdeva2025}. These concerns intensify in advice contexts, where the stance a system takes---validating versus directing, balancing perspectives versus assigning responsibility---shapes what courses of action become thinkable for recipients \citep{vecchione2025artificial}.

\subsection{Vernacular Competence}

The concept that names what the community possesses and aligned models consistently forgo is \textit{vernacular competence}: the locally ratified capacity to convert recognition of harm into directive authorization for action. Top-voted comments routinely perform this conversion in a single move---``This is textbook abuse. Leave.’’ fuses an epistemic assessment with a deontic directive \citep{stevanovic2012deontic}. Through accumulated vote-selection, the community has developed a standard for what counts as advice in the full sense: utterances that not only interpret a situation but authorize a course of action with locally ratified force. Vernacular competence names that standard and the gap its absence produces.

LLM outputs make this standard visible by contrast. The comparison is methodologically useful because it defamiliarizes the community's own standards of adequacy \citep{shklovsky1917art, iliadis2016critical}: by introducing outputs calibrated to portable advisory defaults rather than subreddit-specific criteria, it renders visible what community participation ordinarily leaves unspoken. This works because LLMs function as portable infrastructures with relatively stable advisory defaults across contexts. That stability is what makes the gap analytically productive rather than merely contrastive: it surfaces a recurring pattern in how these systems handle the move from harm recognition to action authorization.

The comparison also makes legible the community’s own taken-for-granted assumptions. On r/relationship\_advice, the conviction that controlling behavior warrants exit operates as a self-evident baseline, presupposing financial independence, material alternatives, and a therapeutic vocabulary through which exit is framed as self-care. These conditions do not need to be argued for because they are not experienced as conditions at all---they are simply what the community treats as obvious. Against that background, the models’ defaults become visible as a portable alternative, not a neutral one.
\section{Data, Taxonomy, and Measurement}
\label{sec:methods}

The article's analysis proceeds in three stages: constructing a topic taxonomy of relationship problems, generating LLM advice responses, and comparing these responses to community-endorsed advice using linguistic metrics. The taxonomy construction stage warrants methodological attention in its own right. Following Bowker and Star's framework for classification as infrastructure \citep{bowker1999sorting}, the categories developed here are treated as analytic choices that encode assumptions about where boundaries fall and what differences matter. How posts are sorted shapes what divergence becomes visible---a point that bears on the interpretation of results in Section~\ref{sec:results}.

\subsection{Corpus, Baseline, and Ethics}

Posts and comments were collected from r/relationship\_advice spanning October 2025 through January 2026, yielding 32,630 posts and 546,521 comments after quality filtering. For the primary comparison, the analysis is restricted to posts with at least one substantive, community-endorsed response: a top-level comment from someone other than the original poster with score $\geq 5$ and length $\geq 200$ characters ($n = 11{,}565$). For each eligible post, LLM-generated advice is compared against the highest-voted qualifying human comment.

R/relationship\_advice is selected as the community baseline precisely because of its structural properties. The community satisfies the methodological conditions for defamiliarization outlined in Section~\ref{sec:bg}: concentrated, measurable norms; a distinctive vocabulary; and a prescriptive force that varies systematically across problem types. The upvote economy of Reddit selects for directiveness and moral certainty; the advice baseline constructed here likely represents a more extreme pole than typical human advice from friends, family, or strangers offline. That extremity is the analytic instrument. The goal is to construct a community baseline whose norms are sufficiently concentrated and legible to make divergence from those norms specifiable. A more diffuse or ambivalent community would flatten the contrast; this one sharpens the comparison. The community is treated as a \textit{baseline for measuring divergence}, not as a normative gold standard. Its own assumptions about exit, autonomy, and the legibility of harm are themselves subjected to critical scrutiny in Section~\ref{sec:discussion}. The divergence documented in Section 4 should therefore be read as the distance between aligned LLM defaults and one particularly concentrated normative formation, not as a deficit relative to human advice more broadly.

No attempt is made to identify individual posters; usernames are not retained and play no role in the analysis, which operates primarily at the aggregate level of linguistic metrics. Illustrative examples are drawn only from high-visibility posts (score > 500), where the assumption of public address is strongest and the advice has already entered wide community circulation.

\subsection{Constructing the Problem Taxonomy}

Standard topic models treat topics as latent distributions over vocabulary, finding lexical co-occurrence patterns rather than semantic categories \citep{Grimmer_Stewart_2013}. In this corpus, the limitation of such an approach is acute, since the vocabulary of relationship distress is relatively uniform across problem types and the distinction between infidelity and controlling behavior lives at the level of narrative framing, not word choice. Both LDA and BERTopic, despite parameter variation, produced only two coarse clusters, because the relevant semantic distinctions occur at the narrative level rather than the word level.

Rather than consolidating toward consensus, six models from diverse providers independently categorize posts (procedural details appear in Appendix~\ref{sec:supplementary}). Their disagreement is treated as data. When six models classify the same post, convergence indicates stable conceptual boundaries; divergence indicates contested terrain where categories are not given by the facts but achieved through particular interpretive frameworks. Each model processed posts sequentially, either assigning to an existing category or proposing a new one. The 930 raw labels were consolidated into a 70-category taxonomy using Claude Sonnet 4 to merge semantically similar labels; cross-model agreement on primary topic was 17.8\%, with systematic disagreements concentrated on harm-versus-neutral framings. This contested terrain is analyzed in Section~\ref{sec:results}. 

\subsection{Generating Model Advice}

Advice was generated using four models (Gemini 2.5 Flash Lite, GPT-4.1-nano, DeepSeek v3.2, and Ministral 8B) across all 11,565 eligible posts. These models were selected as cost-efficient, widely deployed representatives of four distinct providers and training pipelines (Google, OpenAI, DeepSeek, and Mistral), enabling comparison across alignment regimes without the prohibitive cost of frontier models. Notably, no Anthropic model was included in advice generation. Although Claude Sonnet 4 was used in the taxonomy consolidation pipeline---a post-hoc merging task whose output does not enter the comparative analysis---including it as an advice generator would create a conflict of interest in which the same model family both structured the comparison and appeared within it. 

Posts are presented to models with minimal scaffolding: only the post body text is provided, omitting titles and avoiding framing such as ``respond as a Reddit commenter.'' What the model chooses to do, whether advising, validating, or questioning, is itself data about its defaults. This follows work that treats prompting as an empirical audit of generative systems' defaults and normative tendencies, using systematic prompt conditions to surface patterned behavior. Temperature was set to 0.7 for advice generation (characterizing typical deployment behavior) and 0.0 for topic classification (requiring consistency). A sensitivity check at temperature 0.2 ($n = 500$) showed minimal change (mean absolute difference 0.02; divergence persisting at $|d| > 0.5$).

\subsection{Operationalizing Advice Style}

LLM and human advice are compared using lexicon-based metrics, avoiding LLM-as-judge evaluation to prevent circularity. These measures are designed to capture robust directional differences at scale rather than fine-grained semantic content. The analysis makes claims about distributional patterns across thousands of responses, not about the deontic structure of
individual utterances. Key constructs include: \textit{epistemic stance}, operationalized through Hyland's booster/hedge distinction and the deontic/epistemic modal contrast \citep{hyland1998hedging, palmer2001mood}; \textit{relationship orientation}, captured through leave/stay language, red-flag terminology, and therapeutic vocabulary; and \textit{permission-granting}, operationalized as a four-level spectrum following Stevanovic and Per\"{a}kyl\"{a}'s framework for deontic authority transfer \citep{stevanovic2012deontic}: explicit authorization (``you can leave''), conditional authorization (``if he doesn't change, consider leaving''), values affirmation (``you deserve better''), and empathy/validation (``that sounds really hard''). The first two levels grant or approach granting permission to act; the latter two validate feelings without authorizing action. This gradient operationalization addresses construct validity concerns about binary distinctions while preserving the core finding: human advice shows a 5:1 ratio of action-oriented to validation language, while LLM advice inverts this to approximately 1:4. The analysis also measures partner-alignment prompts (``understand their perspective'') and analysis framing (``let's break down''), which index attention direction and advisory stance. Full lexicons appear in Appendix~\ref{sec:supplementary}.

These lexicons are designed as directional proxies for large-scale comparison rather than as fine-grained semantic instruments. To assess whether they recover the broad contrasts central to the argument, two independent coders compared 50 matched human--LLM response pairs on certainty, leave-orientation, and therapeutic framing. When both coders judged that a clear difference existed, they agreed on direction in 96\% of certainty comparisons (24/25 cases), 100\% of leave-orientation comparisons (20/20 cases), and 100\% of therapeutic comparisons (44/44 cases). Disagreements were primarily about whether the difference was large enough to call---one coder marking ``equal'' or ``not applicable'' where the other saw a slight advantage---rather than about which response exhibited more of the property. The near-absence of directional disagreement (only one case where coders said opposite directions) supports the paper's claim that the lexicons recover robust distributional differences in advisory style.

\subsection{Identifying High-Consensus Cases}

To test whether LLM divergence concentrates where human consensus is strongest, the analysis identifies posts where the community has reached clear agreement. Thread-level distributions are used for \textit{selection}: a post qualifies as high-consensus when at least 70\% of its qualifying human comments exhibit leave-oriented language ($n = 153$ posts). This threshold captures clear abuse patterns, discovered infidelity, and safety threats where the community's exit-orientation is most concentrated. For \textit{comparison}, consistency is maintained with the primary analysis by measuring LLM advice against the top comment baseline. On high-consensus posts, the top comment almost invariably reflects the thread's exit orientation; by definition, the voting economy surfaces the directive response when consensus exists. This design tests whether LLM caution scales with situational clarity (as calibrated systems would) or applies uniformly regardless of how certain the community is.

Robustness checks varying the consensus threshold from 50\% to 90\% show a monotonic pattern: as threshold increases, the human-LLM gap widens (0.17 at 50\%, 0.22 at 60\%, 0.30 at 70\%, 0.34 at 80--90\%). The divergence is not an artifact of threshold selection. An alternative operationalization was also considered based on score concentration (whether the top comment dominates thread visibility) but found that concentration alone does not predict divergence. The gap emerges specifically where high concentration coincides with directional consensus: visibility amplifies a settled community position that LLMs do not echo.
\section{Patterns of Divergence Between Community and Model Advice}
\label{sec:results}

This section follows the breakdown between recognition and authorization across two stages: how models classify situations, and how they advise within them. The key question is whether reluctance to recommend action begins with uncertainty about what kind of situation a post describes, or persists even when that uncertainty falls away. Across both analyses, r/relationship\_advice serves as a contrastive baseline: a concentrated, vote-ratified community whose prescriptive clarity makes divergence measurable.

\subsection{Topic Assignment as Moral Perception}

Model disagreement begins at the level of perception. When four models classify the same 11,565 posts into a shared 70-category taxonomy, they reveal different thresholds for what counts as harm. Only 17.8\% of posts receive unanimous agreement; on the remaining 82.2\%, the same case is read differently across models, with some highlighting harm and others reframing it as ordinary friction. If caution in advice were mainly a function of perceptual uncertainty, then agreement on harm classification should narrow the gap in exit recommendations. The analysis below shows it does not.

Consider a 19-year-old woman whose boyfriend insists that spending time with his family every weekend is ``non-negotiable,'' despite knowing she has severe anxiety about large groups. Of four models, only GPT-4.1-nano classifies this as \textit{boundary\_violations}, a harm frame. The other three assign neutral labels: Ministral sees \textit{relationship\_effort\_imbalance}; DeepSeek sees \textit{family\_disapproval\_partner}; Gemini sees \textit{family\_conflict\_dynamics}. The top-voted community response (score: 33), meanwhile, is unambiguous:

\begin{quote}
``Girl run!!! If he can't and won't even consider understanding anxiety than he's not the one. If he's guilt tripping you... that's a huge red flag.''
\end{quote}

The community's voting economy surfaces a harm reading where models remain split. The guilt-tripping and dismissal of anxiety register as a recognizable pattern for regular participants in a forum that has encountered thousands of similar posts, but not to models, which given the task to classify the issue vary in their recognition of harm. This is the community's concentrated interpretive template at work. 

The crucial finding concerns what happens when models' perceptions converge with the community's implicit harm reading. The analysis identified 891 posts where all four models unanimously assigned harm-related categories---cases where perceptual uncertainty cannot explain divergence, since the models agree with each other and implicitly with the community's harm reading. On these posts, community advice is strongly exit-oriented (leave ratio 0.64). If model caution tracked situational ambiguity, convergence on harm classification should narrow the advice gap. Instead, LLM leave ratios on unanimous-harm posts remain at 0.39---roughly 60\% of the community rate, and barely above the rate on posts where no model perceives harm at all (0.28). Models also deploy nine times more hedging language and twenty-eight times more therapeutic framing on these unanimous-harm posts than the community does. The ceiling on directive judgment is not lifted by perceptual agreement.

This finding shows the comparison operating at multiple levels simultaneously. At the level of perception, models' disagreements reveal where conceptual boundaries are contested. But even where perception converges, prescription diverges. The pattern is not a failure of perception but a governed constraint on prescription, operating regardless of how clearly the situation has been identified.

\subsection{Recognition Without Prescription}

Turning from categorization to the advice itself: four models (Gemini 2.5 Flash Lite, DeepSeek v3.2, Ministral 8B, GPT-4.1-nano) generated advice on 11,565 posts, each paired with its highest-scoring community response. LLMs deploy red-flag terminology at rates comparable to community responses, but do not prescribe the solutions the community endorses. Table~\ref{tab:metrics_comparison} summarizes the divergence across six dimensions; the most consequential patterns follow. 

\begin{table}[t]
\centering
\caption{Advice metrics across top-voted r/relationship\_advice comments (``Reddit'') and four LLMs. Higher certainty, modal, and leave ratios indicate more prescriptive advice; higher sentiment and therapy density indicate more uniformly positive, therapeutically-framed responses. Action:validation ratio reflects the four-level permission spectrum described in methods.}
\label{tab:metrics_comparison}
\small
\begin{tabular}{lccccc}
\toprule
\textbf{Metric} & \textbf{Reddit} & \textbf{Gemini} & \textbf{DeepSeek} & \textbf{Ministral} & \textbf{GPT-4.1-nano} \\
\midrule
Certainty ratio & 0.42 & 0.26 & 0.28 & 0.16 & 0.15 \\
Deontic modal ratio & 0.40 & 0.18 & 0.20 & 0.12 & 0.09 \\
Leave ratio & 0.58 & 0.37 & 0.40 & 0.20 & 0.22 \\
Sentiment & 0.13 & 0.81 & 0.63 & 0.96 & 0.97 \\
Therapy (per 1k tokens) & 2.4 & 4.8 & 4.7 & 9.5 & 11.0 \\
Action:validation ratio & 5:1 & 0.3:1 & 0.4:1 & 0.2:1 & 0.2:1 \\
\bottomrule
\end{tabular}
\end{table}

The first divergence concerns \textit{hedging}. Community advice shows a slight preference for confident assertions (``you need to leave'') over hedges (``it might be worth considering''). LLMs systematically hedge regardless of situational clarity, favoring epistemic modals (``might,'' ``could'') that acknowledge uncertainty even when the situation does not warrant it. LLMs hedge even in cases the community treats as requiring directive judgment.

The second concerns \textit{exit recommendations}. Where the community recommends leaving, LLMs recommend ``communication'' and ``working through'' problems. On a post about a partner who monitors the advice-seeker's location and restricts her friendships, the top community response opens: ``This is textbook abuse. Leave.'' The LLM response to the same post: ``Have you tried having an open conversation about how his behavior makes you feel?'' The community's willingness to name abuse and prescribe exit is precisely what LLMs do not reproduce. The divergence marks a shift from directive judgment to communicative process.

\begin{figure}[t]
\centering
\includegraphics[width=0.7\textwidth]{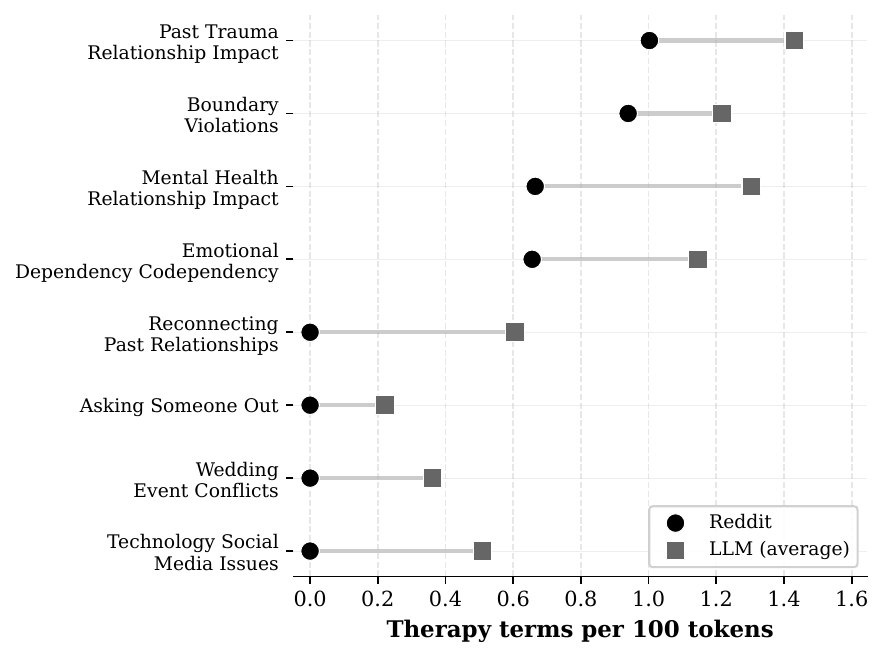}
\caption{Therapeutic language density across eight topics. Reddit (r/relationship\_advice) deployment varies widely (black circles spread from near-zero to elevated), while LLMs maintain uniformly high levels regardless of topic (gray squares). Lines connect values for each topic.}
\label{fig:therapy_distribution}
\end{figure}

The third concerns \textit{therapeutic framing}. LLMs deploy therapy and self-care language (``boundaries,'' ``self-worth,'' ``healing'') at elevated rates across all topics. Yet the difference goes beyond frequency: the coefficient of variation across topics shows that community responses concentrate therapy language selectively (CV = 0.99) while LLMs distribute it uniformly (CV = 0.39). On a post about splitting rent with a higher-earning partner, the community response is transactional: ``Pro-rate it by income. 60/40 if he makes 50\% more.'' The LLM response: ``It's important to honor your own boundaries around financial security while remaining open to understanding his perspective on shared expenses.'' Therapeutic framing appears even where the community judges it inapposite. On practical topics, the community rarely uses therapeutic framing while LLMs deploy it at elevated rates; on emotional topics, both use therapy language, but community responses swing from near-zero to high while LLMs maintain an elevated baseline throughout (Figure~\ref{fig:therapy_distribution}). The community's variation reveals a discriminatory capacity: being able to judge when therapeutic framing is warranted---by its own norms---and when it is not. This capacity is invisible from within the community itself. Alignment has produced a system that cannot \textit{not} therapeutize: one that does not recognize situations where therapeutic language is inapposite by the community's measure, even if other standards might judge it differently.

This flattening extends to affect: LLMs maintain relentlessly positive sentiment (mean 0.84) where the community remains near-neutral (0.13), substituting emotional validation for measured judgment. Where community responses to a difficult situation offer blunt assessment (``He doesn't respect you''), LLMs cushion the same assessment in affirmation (``It's completely understandable that you're feeling hurt, and your feelings are absolutely valid''). The community’s near-neutral sentiment is itself a product of its particular moral formation: bluntness is locally ratified as part of what advice is supposed to do. Analytically, the contrast shows that positivity is not incidental style but part of the model’s advisory default, cushioning judgment in ways that preserve recognition of harm while attenuating permission for action.

\subsection{The Rhetoric of Permission}

The sharpest functional divergence between community and LLM advice emerges in how they grant permission. Following Stevanovic and Per\"{a}kyl\"{a}'s framework for deontic authority in interaction, the analysis distinguishes permission that authorizes concrete action from permission that asserts abstract entitlement \citep{stevanovic2012deontic}. Community advice grants explicit permission to act: ``you can leave,'' ``you don't have to stay.'' These formulations transfer deontic authority, authorizing action the advice-seeker may have felt was forbidden. LLM advice instead asserts abstract entitlements: ``you deserve a partner who respects you,'' ``everyone deserves honesty.'' These validate feelings but do not authorize action; they describe what should be true without granting permission to make it true.

The quantitative pattern is striking. Across a four-level spectrum (explicit authorization, conditional authorization, values affirmation, and empathy/validation), community advice shows a 5:1 ratio of action-oriented language (the first two levels) to validation language (the latter two). LLM advice inverts this to approximately 1:4. The inversion is driven primarily by the final level: LLMs deploy empathy and validation phrases (``that sounds really hard,'' ``I'm sorry you're going through this'') at over 100 times the community rate. In Stevanovic and Per\"{a}kyl\"{a}'s terms, community advice transfers deontic authority; LLM advice lingers at acknowledgment. The difference matters because advice-seekers often already know what they ``deserve''; what they lack---by the community's logic, and by the deontic account this article follows---is permission to act on that knowledge.

A close reading illustrates the gap. A woman posts that her husband monitors her phone, criticizes her friendships, and has started timing how long she takes at the grocery store. The top community response (score: 247) opens: ``This is controlling behavior. You are allowed to have friends. You are allowed to take your time. You do not need his permission to live your life.'' Three moves in quick succession: a diagnostic label that names the behavior as a category (``controlling behavior''), followed by two deontic assertions framed as permissions (``you are allowed''), culminating in an explicit revocation of his authority (``you do not need his permission''). The response closes: ``Start making an exit plan.'' This is a script proposal: a concrete, sequenced action the poster can take.

GPT-4.1-nano's response to the same post opens differently: ``It sounds like you're navigating a really challenging dynamic in your relationship, and it's completely understandable to feel frustrated and constrained.'' Agency is hers, and the problem is relational rather than attributable. ``Completely understandable'' validates the feeling; ``frustrated and constrained'' names emotions rather than behaviors. The response continues with suggestions to ``have an open conversation about boundaries'' and ``consider couples counseling.'' Where the community response grants permission to exit, the LLM response presupposes the relationship will continue and positions the advice-seeker as responsible for managing it. Neither stance is self-evidently correct; the community's exit-orientation reflects specific material and cultural presuppositions that may not hold universally. What the comparison yields is the shape of the divergence, not a verdict on which advice is better.

A second pattern compounds this permission gap. LLMs do not merely refrain from authorizing action, but actively redirect attention toward the partner. This is particularly evident in \textit{partner-alignment prompts}, using phrases such as ``understand his perspective,'' ``consider her feelings,'' ``try to see where they're coming from.'' Community responses show a near-zero rate (0.01 per 1,000 characters). LLM rates are dramatically elevated: Ministral at 32 times the community rate, GPT-4.1-nano at 10 times.

This distinction clarifies the purpose of advice as a speech act. Conversation analytic research has long established that advice-giving is a deontic act: it assumes or creates an asymmetry in which the advisor positions themselves as knowing what the recipient should do. Emmison et al.'s analysis of counseling interactions identifies \textit{script proposals}: formulations that put specific words and actions into the client's repertoire \citep{emmison2011script}. Both the community and LLMs provide scripts---but they are scripts for different actions. The community's ``exit plan'' redistributes obligation; it tells the advice-seeker not only what to do but how to think about what she owes. LLM scripts authorize communication rather than action. The deontic asymmetry differs: one authorizes unilateral action, the other collaborative process. Whether unilateral action is the right call depends on facts the community treats as settled and that LLMs treat as open---which is precisely what the divergence surfaces.

Consider the anxiety-family conflict post. The community response attributes agency: ``Girl, you are an \textit{attorney} now. You have OPTIONS.'' GPT-4.1-nano responds: ``Acknowledge Her Perspective... Try to understand her point of view.'' This advice is not neutral relative to the community's norms. Many people in controlling relationships have already been told they are overreacting, that their partner ``means well.'' The community has encountered thousands of such posts and developed a concentrated interpretive template in which redirecting attention toward the person causing harm replicates the controlling dynamic itself. Against that accumulated template, the model's default becomes visible as structurally consonant with the pattern the advice-seeker is trying to escape---not wrong by some universal standard, but legible as divergence against this particular baseline.

\subsection{Divergence on High-Consensus Cases}

The sharpest test of the comparison method is whether LLMs diverge even on cases where community consensus is overwhelming. 153 posts were identified where at least 70\% of qualifying community comments exhibited leave-oriented language: clear abuse patterns, discovered infidelity, or safety threats. These are ``obvious'' cases from the community's perspective---not necessarily obvious in any absolute sense, but cases where the community's interpretive templates have produced their strongest convergence, and where that convergence can be measured against LLM outputs.

When LLM advice is compared against the top comments on these high-consensus posts, the divergence is stark: the top comments show a leave ratio of 0.87, reflecting the thread-level consensus that elevated them.

\begin{figure}[t]
\centering
\includegraphics[width=0.7\textwidth]{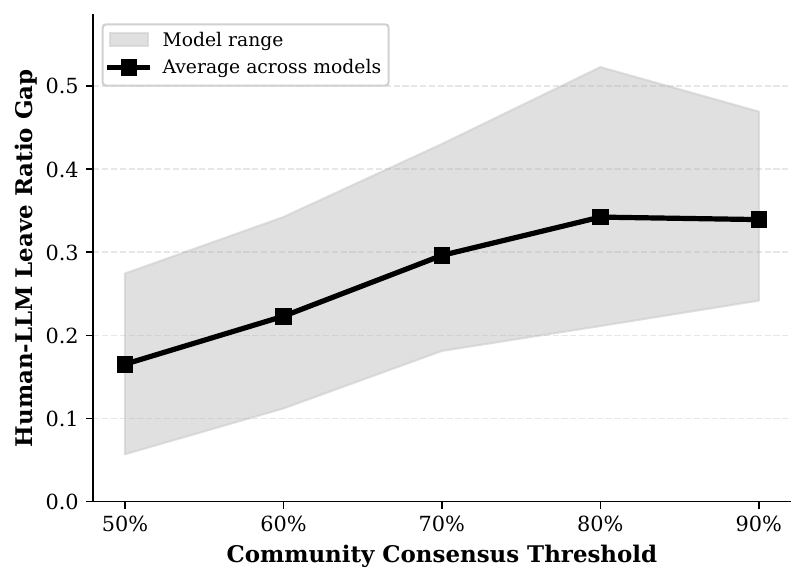}
\caption{
Reddit–LLM divergence as a function of thread consensus. LLMs diverge most from r/relationship\_advice precisely where thread agreement is strongest, indicating uniform suppression rather than calibrated uncertainty. Shaded region shows range across four models.}
\label{fig:consensus_divergence}
\end{figure}

LLM leave ratios remain far lower: Gemini 0.57, DeepSeek 0.56, Ministral 0.34, GPT-4.1-nano 0.37. The divergence is \textit{largest} precisely where community consensus is strongest (Figure~\ref{fig:consensus_divergence}). As the consensus threshold increases from 50\% to 90\%, the gap widens monotonically---the inverse of what a calibrated system would produce if it were tracking the community's confidence. LLMs do not simply diverge from community advice on ambiguous cases; they diverge most dramatically on the cases the community considers least ambiguous.

Table~\ref{tab:qualitative_example} illustrates with a post about a manipulative older partner. The community response leads with six words: ``Break up with him. He is manipulative.'' GPT-4.1-nano leads with thirty-seven words of validation before addressing the situation. Both responses ultimately identify similar concerns; only one opens by telling the advice-seeker what to do.

\begin{table}[t]
\centering
\caption{Qualitative comparison on a high-consensus case. Reddit advice is direct and prescriptive; LLM advice validates feelings while hedging on prescription. The Reddit response exemplifies the subreddit's concentrated prescriptive norm, not a universal ideal of good advice.}
\label{tab:qualitative_example}
\small
\begin{tabular}{p{0.15\textwidth}p{0.75\textwidth}}
\toprule
\textbf{Source} & \textbf{Response (excerpt)} \\
\midrule
Reddit 

(score: 128) & ``Break up with him. He is manipulative. His visa is his responsibility, and at his age he knows that. He was an adult man who dated a teenager and he has planned things so you feel you can't leave him. You can and should. He's able to secure his own future. You deserve your own, away from him.'' \\
\addlinespace
GPT-4.1-nano & ``Thank you for sharing such a detailed and honest reflection on your situation. It's clear that you're weighing a complex set of emotional and practical considerations, and it's understandable that you're feeling conflicted. Regarding the relationship and personal feelings: It's normal for couples to have disagreements and differences, but when core values (such as views on politics, marriage, and raising children) are fundamentally incompatible, it can lead to ongoing dissatisfaction...'' \\
\bottomrule
\end{tabular}
\end{table}

The community response also performs a specific diagnostic: ``He was an adult man who dated a teenager and he has planned things so you feel you can't leave him.'' The comment names the age-gap dynamic and attributes intentionality to the partner's behavior. Crucially, the response provides a script proposal: ``His visa is his responsibility'' is a specific formulation the advice-seeker can adopt and deploy. The LLM offers no such script, treating the situation as a generic values conflict rather than a recognized pattern requiring concrete action. The community's script reflects its accumulated interpretive template for this class of situation---one that treats the age-gap-plus-dependency structure as categorically resolved, where the LLM treats it as open.

Beyond situational clarity, high consensus posts on r/relationship\_advice indicate conformity to the community's existing interpretive templates. Controlling behavior alone accounts for 28\% of high-consensus posts but only 8\% of the full corpus; adding infidelity and explicit breakup considerations brings the total to 48\% of high-consensus posts versus 24\% of the corpus. Categories with less template-conforming narratives, such as family conflict dynamics, emotional dependency, or cultural differences, appear in high-consensus posts at only 0.2--0.4 times their corpus rate. In other words, the community is most certain where its interpretive templates are most developed, and not necessarily where situations are most severe. Posts that fit the controlling-behavior or infidelity scripts reach consensus; posts involving entangled responsibility or contextual complexity do not. The widening gap in Figure~\ref{fig:consensus_divergence} therefore reflects two phenomena simultaneously: LLM suppression of directive judgment, and the community's increasing interpretive certainty on posts that most closely match its existing scripts. Defamiliarization works in both directions: the LLM's flat caution reveals the community's prescriptive force, but the concentration of consensus within template-ready categories reveals that force as categorical rather than deliberative.

\subsection{Model Differences}

\begin{figure}[t]
\centering
\includegraphics[width=0.85\textwidth]{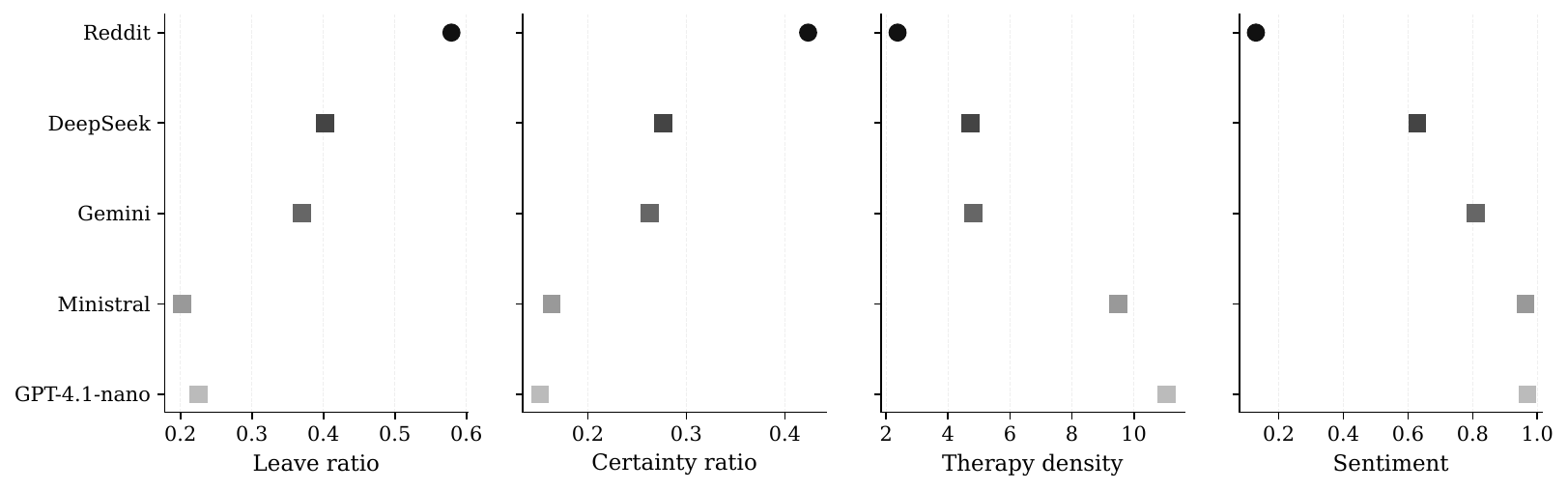}
\caption{Top-voted r/relationship\_advice comments (``Reddit'') and four LLMs compared on leave ratio, certainty ratio, therapy density, and sentiment. DeepSeek is consistently closest to Reddit across metrics while GPT-4.1-nano diverges most; yet no model matches Reddit on leave or certainty, and all exceed Reddit therapy density.}
\label{fig:model_comparison}
\end{figure}

Across the four models, a consistent pattern emerges: GPT-4.1-nano shows the largest divergence from the community baseline across every metric, while DeepSeek consistently falls closest to it (Table~\ref{tab:metrics_comparison}; Figure~\ref{fig:model_comparison}). Gemini and Ministral occupy intermediate positions. Notably, the spread \textit{among} models is itself substantial---DeepSeek's leave ratio (0.40) is nearly twice that of GPT-4.1-nano (0.22)---yet even the closest model remains well below the community baseline (0.58). Inter-model variance and community-model divergence are both real; the former illustrates differences in alignment intensity across training pipelines, while the latter is the primary finding: no model, including DeepSeek, reaches the community's prescriptive floor.

A qualitative comparison on the same post is illuminating. A woman describes a boyfriend who expects her to cook, clean, and commute 70 miles to his apartment, who dismisses her contributions, and who tells her ``I could just hire a chef'' when she falls short of his expectations. DeepSeek opens with a verdict: ``This man is waving more red flags than a communist parade\ldots This man views relationships as a spreadsheet where he tallies `contributions' and expects you to compensate for his financial input with domestic labor, emotional labor, and sex. That's not partnership---it's servitude with benefits.'' It closes with a two-word script proposal: ``RUN.'' GPT-4.1-nano, given the same post, opens: ``Thank you for sharing your story with honesty and detail. It sounds like you're navigating a complex and emotionally taxing situation.'' It identifies the same dynamics of unequal contribution, control, manipulation, but it embeds them in an analytic framework (``Key Issues and Dynamics,'' seven numbered subsections). It closes not with a prescription but with an invitation to reflect: ``reconsidering the relationship altogether.'' The difference is not in diagnostic capacity but in deontic stance---and DeepSeek's closeness to the community's norm does not make its stance correct, only more congruent with this particular moral formation.

A supplementary persona prompting test ($n = 500$) helps locate where these differences originate. When prompted to ``respond as a commenter on r/relationship\_advice would,'' the models narrow the gap on hedging (48\% reduction) and exit recommendations (38\%). But the prompt \textit{widens} the permission gap by 77\%: models become even more validation-oriented when asked to mimic the community. The effect is model-specific and revealing. DeepSeek, persona-prompted, produces recognizably community-style responses (``DUMP HER\ldots Six years is way too long to wait for someone to decide if they want to commit to you''). GPT-4.1-nano, given the same instruction, continues to hedge (``I totally get where you're coming from\ldots it sounds like you've been patient and understanding''). The persona prompt surfaces the community's register in its directiveness, its certainty, and its willingness to assign blame. But it does not lead to a deontic transfer that would make the register functional. Action-authorization is precisely what alignment suppresses most stubbornly, and some models have it suppressed at a level that persona prompting cannot reach.

DeepSeek’s relative closeness to community norms carries a further implication. It is more aligned with this subreddit’s prescriptive style than GPT-4.1-nano, which weakens any simple expectation that provider provenance or broad cultural proximity should predict fit with community judgment. One plausible explanation is that models differ in the extent to which alignment and instruction-tuning push them toward therapeutic neutrality, hedging, and non-directiveness by default. The present design cannot isolate the source of that ordering, whether alignment, training data, or model architecture, but the persona test weighs against a pure incapacity account. A model that can produce community-style advice when prompted yet does not do so by default is one whose advisory stance is being shaped at the level of default behavior rather than raw expressive limits.

\section{Discussion: Alignment, Vernacular Competence, and the Standardization of Judgment}
\label{sec:discussion}

The pattern that emerges across the analysis is a specific normative stance that grants abstract entitlement through therapeutic vocabulary without concrete permission, applied regardless of situational clarity. On r/relationship\_advice, therapeutic concepts function as diagnostic tools that authorize prescriptive conclusions: for instance, identifying ``gaslighting'' warrants recommending people leave their relationships. LLMs deploy the same concepts as a uniform register, present everywhere but discriminating nowhere, so that the diagnostic function is severed from its prescriptive consequence.

\subsection{A Governed Ceiling on Directive Judgment}

The most striking finding is structural: on the 153 posts where at least 70\% of community comments recommend leaving, models do not converge with the community; the divergence widens. This is the inverse of what a system calibrated to situational clarity would produce. If LLM caution tracked genuine uncertainty, models would hedge most on ambiguous cases and converge with community responses on clear ones. Instead, the models apply a ceiling on prescriptive confidence regardless of situational clarity, creating an inverse relationship between community certainty and model directiveness. The ceiling does not reflect judgment about particular cases; it reflects a prior about what kind of speech is permissible. Safety alignment is the most plausible account of this prior, though the design cannot rule out contributions from training-data composition and assistant-style instruction tuning.

This has a consequence for understanding safety alignment. The result is not ``bad advice'' but something more precisely characterizable: a
subject who feels validated but is not authorized to act, encouraged to manage emotional responses rather than change a situation. Where the community's therapeutic vocabulary leads to action (``this is 
gaslighting, therefore leave''), LLM therapeutic vocabulary terminates in self-management---what Illouz calls the empowering potential of therapeutic culture \citep{illouz2008saving}, inverted into a form of 
responsibilization that locates structural harm in the victim's capacity for emotional regulation. What alignment yields, more precisely, is \emph{techne} without \emph{phronesis}. 

Vernacular competence in this community, then, is not primarily about emotional attunement or diagnostic accuracy---LLMs perform adequately on both. It is about what Stevanovic and Peräkylä call deontic authority transfer: the willingness to tell someone what to do, and to provide the specific formulation that makes doing it thinkable. The patterns documented here are consistent with this framework: models less often convert diagnostic recognition into explicitly action-guiding language, and when they do, they hedge the prescription in ways the community does not. Lu's politeness research confirms this has evaluative force: generic advice violates ``sociality obligations,'' the expectation that advisers provide tailored, enactable guidance rather than blanket statements \citep{lu2025politeness}. What appears in these outputs is not just suppressed judgment but a recurring substitution: context-specific prescription gives way to a portable advisory style based on caution, validation, and broadly acceptable speech.

The method developed here makes this normative stance empirically characterizable in ways that existing evaluation approaches often do not. Red-teaming tests whether models produce harmful outputs; benchmarks assess helpfulness against generic standards. Neither specifies the structure of alignment's moral economy, the particular configuration of therapeutic uniformity, prescriptive ceilings, and attention-redirection documented here. 

This is not to say that the community's exit-oriented advice is ``unsafe''; it is a different theory of what advice should do. Vicsek et al.\ find that models struggle when cultural sensitivity conflicts with human rights principles in cross-cultural contexts \citep{vicsek2025crosscultural}. The findings reveal a parallel tension operating within Western contexts: safety alignment encodes a particular cultural position (therapeutic neutrality) that conflicts with other defensible positions, including directive intervention in situations the community recognizes as dangerous.

It is worth stating plainly that defamiliarization, if it is to be more than a tool for criticizing one side, must work in both directions. The goal is not to recommend that LLMs behave more like r/relationship\_advice, but to characterize what each system assumes and what each makes visible about the other. The community's directiveness rests on assumptions that deserve scrutiny: directive exit advice from strangers, based on one-sided accounts, carries real risks. The community cannot know what the poster omitted, cannot follow up to learn whether its advice helped, and cannot verify that exit is materially possible. High consensus may reflect template-matching to familiar scripts as much as situational clarity. LLM caution, whatever its origins, does not presume these conditions are met. 

But the reverse is also true: the community's capacity to recognize coercive control patterns, isolation tactics, and grooming represents accumulated practical knowledge that therapeutic neutrality cannot access. The analysis demonstrates that aligned LLMs occupy a  consistent normative position, forgoing deontic authorization in ways that do not vary with situational clarity. The question is not which moral economy is correct, but what it means that the threshold between recognizing harm and authorizing action is being set in advance by design.

The repertoire documented in Section~\ref{sec:results} around therapy, communication, reflection, and self-work encodes an implicit theory of what relationship problems are. If problems are located in the advice-seeker's pattern-recognition or self-worth rather than in another person's behavior, then solutions are correspondingly internal: grow, reflect, communicate better. The community's contrasting repertoire locates some problems in the other person (``this is textbook abuse,'' ``he planned things so you can't leave'' etc.) and correspondingly prescribes external action: exit, not self-work. Both are theories of problems, not merely response styles. Alignment assumes problems are communicatively and therapeutically tractable; the community assumes some are structural, requiring action rather than dialogue.

\subsection{Implications for Evaluation}

Current evaluation frameworks treat advice quality as optimizable against a single standard---typically helpfulness as judged by users or other LLMs. Benchmarks assess whether advice is helpful, harmless, and honest, but not whether it grants permission to act. The operationalization of deontic authority transfer developed here suggests this is a measurable property rather than an ineffable quality. Benchmarks that reward only validation will systematically prefer advice that does not provide the authorization situated communities do. The fact that LLM hedging concentrates on topics where the community's exit orientation is strongest, such as abuse, safety threats, and controlling behavior, also raises questions about whether harms of omission deserve consideration alongside the harms that current alignment research targets.

As LLMs become prevalent in advice-seeking contexts, their systematic divergence from community-specific practices could gradually reshape those practices. The encounter between AI defaults and platform cultures is not a one-time comparison but an ongoing interaction whose dynamics warrant longitudinal study. Evidence from other consequential domains shows that LLM-mediated drafting can alter institutional outcomes (e.g., public comments), suggesting that repeated exposure to model defaults can re-pattern social practices over time \citep{arsenault2026whose}. The therapeutic flattening documented here thus provides specificity to what Kleinberg and Raghavan describe as the risk of algorithmic monoculture \cite{kleinberg2021algorithmic}. If advice-seekers increasingly encounter the same response structures, moral diversity may erode not through any singularly harmful response but through the cumulative displacement of situated judgment by averaged defaults.

\subsection{Limitations}

Several limitations qualify the interpretation of these findings and clarify the scope of the claims advanced here. The four models used here are cost-optimized rather than frontier systems; the patterns observed could reflect reduced capacity rather than safety alignment, and replication with frontier models would 
strengthen mechanistic claims. The persona prompting results reported in Section 4.5 weigh against a purely capacity-based account: a model that produces community-style responses when explicitly prompted but not by default is a model whose defaults have been shaped by something other than incapacity. This does not rule out capacity effects, but it shifts the burden toward alignment-related explanations. Reagrding model choice, it is possible that Anthropic models would show different patterns, and the findings cannot be assumed to generalize to systems with explicitly constitutionalist alignment approaches.

The linguistic metrics used here trade semantic precision for scalability. Pairwise validation on 50 human–LLM response pairs supports their directional adequacy for the present analysis: human responses were judged more certain in 72\% of cases, more leave-oriented in 76\%, and LLM responses more therapeutically framed in 96\%. At the same time, these checks do not establish that the lexicons capture the full semantic or pragmatic structure of individual utterances. Inter-coder reliability was uneven, underscoring the difficulty of drawing sharp boundaries in overlapping and highly skewed advice categories. The measures are therefore best understood as coarse instruments for identifying large-scale directional contrasts in advisory style, and not as exhaustive accounts of how judgment, authorization, or therapeutic framing operate in every response.

The analysis focuses on one subreddit representing a specific slice of American online discourse. The posts analyzed reflect multiple selection processes, so the findings characterize \textit{endorsed community advice}, not representative human advice. A related limitation concerns using this community as a baseline at all: r/relationship\_advice's upvote economy actively selects for certainty and directiveness, which means the community's norms around hedging and prescription are themselves distinctive rather than representative of human advice more broadly. The divergence documented here is partly a function of how concentrated these norms are; a community with more diffuse or ambivalent standards would produce smaller gaps and weaker contrast effects. This does not undermine the analysis---the method requires concentrated norms---but it means the findings characterize the LLM's distance from a particular extreme, not from typical human advice. Training data averaging cannot be definitively separated from safety alignment as mechanisms, though the finding that divergence widens where uncertainty is lowest is more consistent with a governed constraint on prescription than with simple averaging.
\section{Conclusion}

Instead of treating alignment as an intrinsic property of LLMs---whether models are helpful, harmless, or honest---it is better understood as a situated relationship between model behavior and the social contexts in which it is interpreted and evaluated. A model aligned with the averaged norms of diffuse internet training is misaligned with any concentrated community's specific wisdom. A model trained to avoid prescriptive language is misaligned with contexts where adequate advice includes authorizing action. 

The question for researchers and designers is not whether LLMs are aligned, but who might benefit from that alignment, and who might suffer. For advice-seekers facing clear-cut situations of harm, the consequences of the current study are practical: the permission to act that communities grant and LLMs do not is not a stylistic difference but a functional one. This is consistent with understanding safety alignment as a specific form of algorithmic moral governance, whose practical effect is to prioritize corporate risk-aversion and therapeutic self-management over the situated, directive friction often necessary for escaping harm.

The concept of vernacular competence, and the method of reading AI outputs against concentrated community norms, offer tools for investigating these dynamics beyond the case examined in this article. Any domain where communities have developed situated practices of authorization possesses a vernacular competence that LLM contrast can render visible. What the analysis shows is that assistant-style AI carries a portable normative style that flattens situated modes of judgment, and that this matters most precisely where communities treat directive action as necessary rather than excessive.
\appendix

\section{Supplementary Materials}
\label{sec:supplementary}

\subsection{Advice Generation Prompt}
\label{sec:prompt-advice}

For advice generation, no prompt scaffolding is used. The model receives only the post body text, with no instructions, framing, or persona guidance. This design choice is methodologically motivated: what the model \textit{chooses to do} when presented with a relationship dilemma---advise, validate, question---is itself data about its defaults.

\begin{quote}
\small
\ttfamily
\{post\_body\_text\}
\end{quote}

\textbf{Parameters}: temperature=0.7, max\_tokens=4096

Post titles are omitted to avoid activating genre-specific patterns from Reddit's characteristic title format.

\subsection{Model Access and Traceability}
\label{sec:model-access}

All models were accessed via OpenRouter (\url{https://openrouter.ai}) between January 20--25, 2026. Table~\ref{tab:model_details} provides model identifiers and access details for reproducibility.

\begin{table}[h]
\centering
\small
\caption{Model access details. All models accessed via OpenRouter API.}
\label{tab:model_details}
\begin{tabular}{llll}
\toprule
\textbf{Model (paper)} & \textbf{API identifier} & \textbf{Provider} & \textbf{Access dates} \\
\midrule
Gemini 2.5 Flash Lite & google/gemini-2.5-flash-lite & Google & Jan 20--25, 2026 \\
GPT-4.1-nano & openai/gpt-4.1-nano & OpenAI & Jan 20--25, 2026 \\
DeepSeek v3.2 & deepseek/deepseek-chat-v3-0324 & DeepSeek & Jan 20--25, 2026 \\
Ministral 8B & mistralai/ministral-8b & Mistral AI & Jan 20--25, 2026 \\
\addlinespace
\multicolumn{4}{l}{\textit{Topic discovery only:}} \\
Gemini 2.0 Flash & google/gemini-2.0-flash-001 & Google & Jan 21, 2026 \\
Claude Haiku 4.5 & anthropic/claude-3-5-haiku & Anthropic & Jan 21, 2026 \\
\bottomrule
\end{tabular}
\end{table}

Model versions may change over time; the identifiers above reflect the versions available through OpenRouter during the access period. Access to model weights, training data, or alignment procedures was not available beyond what providers publicly document.

\subsection{Topic Discovery and Assignment}

Topic discovery used six models (Gemini 2.0 Flash, Claude Haiku 4.5, Gemini 2.5 Flash Lite, DeepSeek v3.2, Ministral 8B, GPT-4.1-nano) to independently categorize posts, following \citet{pham2024topicgpt}. Each model processed posts sequentially with the following prompt:

\begin{quote}
\small
\ttfamily
You will receive a post and a set of existing topic categories. Your task is to identify the core topic of this post.

[Existing Topics]\\
\{topics\}

[Instructions]\\
1. Read the post and identify its PRIMARY topic.\\
2. Topic labels must be GENERALIZABLE (not specific to this post).\\
3. If the post fits an existing topic, use that exact topic.\\
4. If no existing topic fits well, propose ONE new topic.

[Post]\\
\{post\_text\}

Respond with JSON: \{"action": "existing" or "new", "topic": "topic\_label"\}
\end{quote}

This design allows the taxonomy to emerge from model judgments rather than being imposed a priori. The 930 raw labels were consolidated into a 70-category taxonomy using Claude Sonnet 4. Topic assignment then used four models to assign 1--3 topics per post from the consolidated taxonomy. All discovery and assignment used temperature=0.0.

\subsection{Lexicon Definitions}
\label{sec:lexicons}

The following lexicons operationalize the linguistic constructs in Section~\ref{sec:methods}. All matching is case-insensitive. Where established lists exist, they are adapted for the advice-giving domain; where no standard lists exist, domain-specific lexicons are developed and validated against human judgment (see below).

\paragraph{Certainty Markers.}
Adapted from Hyland's taxonomy of hedges and boosters in academic discourse \citep{hyland2005metadiscourse}, modified for informal advice-giving (e.g., including ``maybe,'' common in advice but rare in academic prose; excluding meta-discourse verbs like ``argue'' and ``claim'' that mark academic stance rather than interpersonal certainty).
\textit{Hedges}: might, could, may, perhaps, possibly, maybe, probably, seemingly, apparently, potentially, somewhat, fairly, rather; \textit{phrases}: it seems, it appears, I think, I believe, not sure, hard to say.
\textit{Boosters}: clearly, definitely, obviously, absolutely, certainly, always, never, completely, totally, entirely, utterly.
\textit{Certainty ratio}: boosters / (hedges + boosters).

\paragraph{Modal Verbs.}
Following standard classifications of modal meaning \citep{palmer2001mood}.
\textit{Deontic}: should, must, ought, need to, have to.
\textit{Epistemic}: might, could, may, would, can.
\textit{Modal ratio}: deontic / (deontic + epistemic).

\paragraph{Relationship Orientation.}
Domain-specific lexicons developed for this study based on preliminary analysis of r/relationship\_advice discourse patterns.
\textit{Leave words}: leave, break up, divorce, end it, walk away, dump, move on, get out, run, separate.
\textit{Stay words}: stay, work on, communicate, couples therapy, marriage counseling, work through, reconcile, forgive, compromise.
\textit{Leave ratio}: leave / (leave + stay).

\paragraph{Diagnostic Language.}
Domain-specific lexicons capturing the therapeutic vocabulary characteristic of online relationship discourse.
\textit{Red flags}: red flag, toxic, narcissist, gaslighting, manipulat*, abusive, controlling, grooming, love bomb, coercive.
\textit{Therapy words}: boundar*, self-worth, therapy, therapist, self-care, mental health, healing, trauma, self-esteem, codependen*, well-being.

\paragraph{Permission Language.}
Based on the deontic authority framework of \citet{stevanovic2012deontic}, distinguishing formulations that authorize concrete action from those that assert abstract entitlement.
\textit{Action-oriented}: you can leave, you don't have to, you're allowed to, you have every right to, it's okay to, you don't owe.
\textit{Abstract entitlement}: you deserve, everyone deserves, you're worthy of, you merit.

\subsection{Validation}

The lexicon-derived metrics were validated using pairwise comparison, a design suited to the relative nature of the claims. Rather than asking whether absolute scores are accurate, the validation asked whether metrics correctly identify which response in a human--LLM pair exhibits more of a given property. Two independent coders evaluated 50 pairs drawn from leave-relevant topics (controlling behavior, infidelity, breakup considerations), judging for each pair whether the human or LLM response was more certain, more leave-oriented, and more therapeutic. Coders could mark pairs as ``equal'' (no clear difference) or ``skip'' (dimension not applicable to this pair).

The key finding concerns directional agreement: when both coders judged that a clear difference existed, they overwhelmingly agreed on which response exhibited more of the property. For certainty, coders agreed on direction in 96\% of cases where both made a directional judgment (24/25). For leave-orientation, directional agreement was 100\% (20/20). For therapeutic framing, agreement was 100\% (44/44)---both coders consistently identified LLM responses as more therapeutic.

Disagreements were concentrated at the threshold rather than the direction. One coder frequently marked ``equal'' or ``skip'' where the other saw a slight difference favoring human responses. This pattern indicates that coders applied different thresholds for what counts as a meaningful difference, not that they perceived opposite patterns. Only one case across all three dimensions involved true directional disagreement (one coder saying ``human'' while the other said ``LLM''). The near-absence of directional disagreement supports the validity of the lexicon-based measures for recovering the broad contrasts central to this analysis.


\section*{Declaration of conflicting interests}
The author declares no potential conflicts of interest with respect to the research, authorship, and/or publication of this article.

\section*{Funding}
The author received no financial support for the research, authorship, and/or publication of this article.

\section*{Data availability}
The analysis code and processed data supporting this study are available at \url{https://github.com/tomvannuenen/recognition-without-prescription}.
Raw Reddit data were collected via the Reddit API and are subject to Reddit's Terms of Service; the repository includes the post IDs necessary to reconstruct the dataset.

\bibliography{custom}

\end{document}